\documentclass{aa} 
\usepackage{graphicx}
\usepackage{color}
\usepackage{txfonts}
\usepackage{natbib}
\usepackage{hyperref}
\usepackage[normalem]{ulem}
\usepackage{soul}
\newenvironment{myfont}{\fontfamily{lmtt}\selectfont}{\par}
\DeclareTextFontCommand{\textmyfont}{\myfont}
\input epsf.sty

\def\civ{C{\sc iv}}
\def\kms{\,km\,s$^{-1}$}

\def\hb{{\sc{H}}$\beta$\/}

\def\siiv{Si{\sc iv}\/}
\def\oiv{O{\sc iv]}\/}
\def\nv{{N\sc{v}}}
\def\heii{He{\sc{ii}}}
\def\aliii{Al{\sc  iii}}

\def\feii{Fe {\sc{ii}}}
\def\caii{Ca {\sc{ii}}}
\def\rfe{R$_{\rm{FeII}}$}
\def\rcat{R$_{\rm{CaT}}$}
\def\cat{CaT}

\def\mbh{$M\mathrm{_{BH}}$}

\def\mdot{$\dot{\mathcal{M}}$}

\def\RL{$R\mathrm{_{H\beta}}-L_{5100}$}

\def\zsun{Z$_{\odot}$}
\def\msun{M$_{\odot}$}
\def\un{$\log {U} - \log {n_{H}}$}
\def\rblr{$R\mathrm{_{BLR}}$}

\def\hb{H$\beta$}

\def\n{${n_{H}}$}
\def\u{${U}$}

\begin{document}

\title{The CaFe Project: Optical \textmyfont{Fe II} and Near-Infrared \textmyfont{Ca II} triplet emission in active galaxies -- simulated EWs, the co-dependence of cloud sizes and metal content}

\author{Swayamtrupta Panda\inst{1,2}}

\offprints{panda@cft.edu.pl}

\institute{Center for Theoretical Physics, Polish Academy of Sciences, Al. Lotnik\' ow 32/46, 02-668 Warsaw, Poland\\ \and
Nicolaus Copernicus Astronomical Center, Polish Academy of Sciences, Bartycka 18, 00-716 Warsaw, Poland}

\abstract
{} 
{Modelling the low ionization lines (LILs) in active galactic nuclei (AGNs) still faces problems in explaining the observed equivalent widths (EWs) when realistic covering factors are used and the distance of the broad-line region (BLR) from the centre is assumed to be consistent with the reverberation mapping measurements. We re-emphasize this problem and suggest that the BLR ``sees'' a different continuum compared to a distant observer. This change in the continuum reflected in the change in the net bolometric luminosity from the AGN is then able to resolve this problem. \\}
{We carefully examine the optical \feii{} and near-infrared \caii{} triplet (\cat{}) emission strengths, i.e. with respect to H$\beta$ emission, using the photoionization code \textmyfont{CLOUDY} using a range of physical parameters, prominent among which are (a) the ionization parameter (\u{}), (b) the local BLR cloud density (\n{}), (c) the metal content in the BLR cloud, and (d) the cloud column density. Using an incident continuum for \textmyfont{I Zw 1} - a prototypical Type-1 narrow-line Seyfert galaxy, our basic setup is able to recover the line ratios for the optical \feii{} (i.e. \rfe{}) and for the NIR \cat{} (i.e. \rcat{}) in agreement to the observed estimates. Although, the pairs of (\u{},\n{}) that reproduce the conforming line ratios, unfortunately, do not relate to agreeable line EWs. We thus propose a way to mitigate this issue - the low ionization line (LIL) region of the BLR cloud doesn’t see the same continuum seen by a distant observer that is emanated from the accretion disk, rather it sees a filtered version of the original continuum which brings the radial sizes in agreement with the reverberation mapped estimates for the extension of the BLR. This is achieved by scaling the radial distance of the emitting regions from the central continuum source using the photoionization method in correspondence with the reverberation mapping estimates for \textmyfont{I Zw 1}. Taking inspiration from past studies, we suggest that this collimation of the incident continuum can be explained by the anisotropic emission from the accretion disk that modifies the spectral energy distribution (SED), such that the BLR receives a much cooler continuum with a reduced number of line-ionizing photons that allows mitigating the issue with the lines' EWs. \\}
{(1) The assumption of the filtered continuum as the source of BLR irradiation recovers realistic EWs for the LIL species, such as the H$\beta$, \feii{} and \cat{}. However, our study finds that to account for the adequate \rfe{} (\feii{}/H$\beta$ flux ratio) emission, the BLR needs to be selectively overabundant in iron. On the other hand, the \rcat{} (\cat{}/H$\beta$ flux ratio) emission spans a broader range from solar to super-solar metallicities. In all these models the BLR cloud density is found to be consistent with our conclusions from prior works, i.e. \n{} $\sim$ 10$^{12}$ cm$^{-3}$ is required for the sufficient emission of \feii{}, as well as for \cat{}.\\

(2) We extend our modelling to test and confirm the co-dependence between the metallicity and the cloud column density for these two ionic species (\feii{} and \cat{}), further allowing us to constrain the physical parameter space for the emission of these LILs. Adopting the estimates from line ratios that diagnose the metallicity in these gas-rich media that suggest super-solar values ($\gtrsim$ 5-10 \zsun{}), we arrive at cloud columns that are of the order of 10$^{24}$ cm$^{-2}$.\\

(3) Finally, we test the effect of inclusion of a micro-turbulent velocity within the BLR cloud which informs us that the \feii{} emission is positively affected by the inclusion of the microturbulence. An interesting result obtained here is the reduction in the value of the metallicity by up to a factor of 10 for the \rfe{} cases when the microturbulence is invoked, suggesting that microturbulence can act as an apparent metallicity controller for the \feii{}. On the contrary, the \rcat{} cases are rather unaffected by the effect of microturbulence.}
{}

\authorrunning{Panda}
\titlerunning{\textmyfont{Fe II} and \textmyfont{Ca II} emission in the BLR of AGNs}
\maketitle

\section{Introduction} 
\label{sec:intro}

The Broad-line region (BLR) of active galactic nuclei (AGN) consists of two basic components emitting the high ionization lines (HILs) and low ionization lines (LILs), and the two regions have different physical conditions and show different dynamical motion \citep{collin88, 2019A&A...627A..88M}. LIL part comes from a denser region, closer to the mid-plane of the black hole accretion disk system, one that is perpendicular to the black hole's spin axis, and it plays a very important role in two aspects, firstly, in black hole mass measurements using the H$\beta$ and Mg II \citep[][]{czerny2019,2020ApJ...901...55H,2020ApJ...896..146Z,2020ApJ...903...86M, 2020arXiv201212409Z}. The typical values for time-lags reported for the LILs (e.g., H$\beta$ (IP: 13.6 eV) and Mg II (IP: 15.04 eV\footnote{The values for the ionization potential (IP) are taken from \href{https://physics.nist.gov/PhysRefData/ASD/ionEnergy.html}{NIST Atomic Spectra Database Ionization Energies}})) are found to be longer than those shown for the HILs (e.g., C IV (IP: 64.49 eV) and He II (IP: 54.42 eV)) from reverberation mapping studies \citep{peterson_wandel_1999,horne2021}. And secondly, in defining the quasar main sequence, which additionally involves the optical Fe II emission \citep[][]{bg92,sul00,sh14,mar18,panda18b,panda19b}. Here, the optical Fe II emission refers to the 4434-4684 \AA~ blend blue-wards of the H$\beta$ emission line. This definition for the optical Fe II is employed throughout this paper.

Despite its importance and long years of studies, the modelling of the LILs is inherently difficult \citep[e.g.][]{collin-souffrin1986,jol87,kor97}. In the case of Fe II lines this is connected with a huge number of transitions which should be incorporated into the radiative transfer computations \citep[e.g. in \textmyfont{CLOUDY},][]{f17}. In spectral analysis, observational \citep[][]{bg92,vestegaard01} and semi-empirical \citep[][]{veron2004,kovacevic2010,rissmann2012} templates are frequently used.
The difficulty in understanding the \feii{} emission has led us in search of other reliable, simpler ionic species such as \caii{} and O I \citep[][and references therein]{martinez-aldamaetal15} which would originate from the same part of the BLR and have potential to play a similar role in quasar main sequence studies. Here, the \caii{} emission refers to the \textit{Ca II triplet} (\cat{}), i.e., the IR triplet emitting at $\lambda$8498\AA, $\lambda$8542\AA\, and $\lambda$8662\AA. We refer the readers to \citet[][hereafter P20]{pandaetal2020_paper1} for an overview on the issue of \cat{} emission in AGNs and its relevance to the \feii{} emission.

In P20, we compiled an up-to-date catalogue of quasars with spectral measurements of the strengths of the optical \feii{} and NIR \cat{} emission, i.e., for the Fe II, this is the intensity ratio of the optical Fe II blend between 4434-4684 \AA~ measured to \hb{}. This is denoted as \rfe{}. Similarly, for the \cat{}, this refers to the intensity ratio of the \cat{} emission measured to \hb{}. This is denoted as \rcat{}. Our findings in P20 reinforced the existing tight correlation \citep{martinez-aldamaetal15} between the strengths of the aforementioned two ionic species with a higher significance. We also performed a suite of \textmyfont{CLOUDY}\footnote{\href{https://gitlab.nublado.org/cloudy/cloudy/-/wikis/home}{https://gitlab.nublado.org/cloudy/cloudy/-/wikis/home}} photoionisation models to derive the \rfe{}-\rcat{} correlation from a theoretical standpoint with emphasis on the important roles played by the ionization parameter and the local cloud density. We touched upon the effect of metallicity and cloud column density and show their marked contribution to this correlation, albeit qualitatively.

While P20 was devoted to justifying the connection between the optical \feii\ and NIR \cat{}, there we used only the line ratios and did not address the basic problem of the LIL-emitting region which is the reproduction of the observed equivalent widths (EWs) of the lines. The main goal of the present paper is to match the modelled data with the observations in terms of the lines' EWs and flux ratios of these two ions and constrain the relative location of \feii\ and \cat{}, and to determine the metallicity required to optimize the emission strengths of these two ions. Additionally, this paper investigates the effect of the cloud column densities (N$_{\rm{H}}$) on the net emission strengths of the aforementioned ions, which, for a given local mean density of the BLR cloud, estimates the size of the BLR cloud. The treatment of the metallicity and cloud column density is done in a heuristic manner and the obtained inferences are gauged against prior observed measurements for \textmyfont{I Zw 1}.

The paper is organised as follows: In Section \ref{sec:methods}, the photoionisation modelling setup is described keeping aligned with our approach in P20. The novelty of this part of the work lies in (i) an appropriate treatment of the issue of the equivalent widths in terms of the covering factor for the line species; and (ii) a systematic treatment of the metallicity and cloud column density unlike P20, where we assumed only two representative cases for each entity, i.e. Z = 0.2\zsun{} and 5\zsun{} at N$\rm{_{H}} = 10^{24}$ cm$^{-2}$, and, N$\rm{_{H}} = 10^{24.5}$ cm$^{-2}$ and $10^{25}$ cm$^{-2}$ at Z = \zsun{}. In Section \ref{sec:results},  we analyse the results from the photoionization models and check for inconsistency with regards to the line equivalent widths of \hb{}, optical \feii{} and \cat{}, previously noticed in P20, and propose a way to bring the results from photoionization modelling in agreement with the observational estimates. We discuss certain aspects of the results and their implications in Section \ref{sec:discussion}. The key findings from this study are then summarized in Section \ref{sec:conclusion}.

\section{Photoionization simulations with CLOUDY}
\label{sec:methods}

According to the standard photoionization theory, the ratio of the hydrogen ionizing photon density to the total hydrogen density is denoted by the dimensionless ionization parameter \u{}, such that:
\begin{equation}
    U = \frac{Q_{\rm{H}}}{4\pi r^2n_{\rm{H}}c}
\end{equation}
where $Q_{\rm{H}}$ is the number of hydrogen ionizing photons emitted by the central object, $r$ is the separation between the centre of the source of ionizing radiation and the illuminated face of the line-emitting medium, \n{} is the total hydrogen (or mean cloud) density and $c$ is the speed of light. All parameters have cgs units.  
In accordance to P20, we perform a suite of \textmyfont{CLOUDY} \citep[version 17.02,][]{f17} models\footnote{N(\u{}) $\times$ N(\n{}) $\times$ N(Z) = 12$\times$11$\times$5 = 660 models}
by varying the mean cloud density, $10^{10.5} \leq n_H \leq 10^{13}\;(\rm{cm^{-3}}$), the ionization parameter, $-4.25 \leq \log U \leq -1.5$, the metallicity, 0.1\zsun{} $\leq$ Z $\leq$ 10\zsun{}, at a base cloud column density, $N_{\rm{H}} = 10^{24}$ cm$^{-2}$. The cloud column density command allows the user to set the size ($d$) of the line emitting medium given by the relation: $d = \frac{N_{\rm{H}}}{n_{\rm{H}}}$, where $N_{\rm{H}}$ and $n_{\rm{H}}$ have their usual meaning. Other cases of cloud column densities are explored in later sections. 
As in P20, we have used a constant shape for the ionizing continuum one that is appropriate for the nearby NLS1, \textmyfont{I Zw 1} \footnote{The \textmyfont{I Zw 1} ionizing continuum shape is obtained from \href{http://ned.ipac.caltech.edu/byname?objname=I\%20Zw\%201\&hconst=67.8\&omegam=0.308\&omegav=0.692\&wmap=4\&corr_z=1}{NASA/IPAC Extragalactic Database}}. The bolometric luminosity of \textmyfont{I Zw 1} is L$_{\rm{bol}} \sim 4.32\times10^{45}$ erg s$^{-1}$. This is obtained by applying the bolometric correction prescription from \cite{netzer2019}:
\begin{equation}
    k_{\rm{Bol}} = c \times \left(\frac{L_{5100}}{10^{42}}\right)^d
\end{equation}
where, c=40, d=-0.2 and $L_{5100}$ is measured in erg s$^{-1}$. For \textmyfont{I Zw 1}, $\rm{L_{5100}} \sim 3.48\times 10^{44}\;\rm{erg\;s^{-1}}$ \citep{persson1988}. \citet{huang2019} performed the first reverberation mapping campaign for this source and obtained a value for the $L_{5100}$ = 3.19$^{+0.27}_{-0.27}\times 10^{44}\;\rm{erg\;s^{-1}}$ which agrees well with the previous estimate from \citet{persson1988} within 1$\sigma$ uncertainty. Since we also use the \rfe{} and \rcat{} estimates from their study \citep{persson1988}, we incorporate their $L_{5100}$ value for our study.

Compared to the range of \n{} and \u{} explored in P20, both entities are extended by 1 dex to explore possible solutions in a low density-low ionization regime. Compared to P20, we do not impose any limitation on the \un{} space due to the dust sublimation. The model assumes a distribution of cloud densities at various radii from the central illuminating source to mimic the gas distribution around the close vicinity of the active nuclei. The range of metallicity incorporated here is inspired by the works on quasar main sequence containing a distribution of quasars ranging from the low-\rfe{} ``normal'' Seyfert galaxies which can be modelled with sub-solar assumption and the Narrow-line Seyfert galaxies (NLS1s), especially the extreme \feii{} emitters, that are proposed to have super-solar metallicities \cite[][]{laor971zw1,negreteetal12,Marziani2019, 2020arXiv200914177S}. Also, the range of cloud column density used is in agreement with previous works, mainly in \citet{perssonferland89, matsuoka07, matsuoka08, negreteetal12} and further extension shown in P20. The line fluxes, \rfe{} and \rcat{} estimates are extracted from these simulations. 

In the following sections, we analyse the results from the photoionization models and check for inconsistency with regards to the line equivalent widths of \hb{}, optical \feii{} and \cat{}. We then apply a simple \textit{radiation filtering} to the incident continuum to mimic the incoming radiation seen by the BLR cloud which scales down the radial distance of the BLR cloud in agreement with the reverberation mapping results. This filtering also brings the EWs and their corresponding covering factors to be consistent with the observed data. Next, by imposing an additional constraint on the obtained estimates for \rfe{} and \rcat{} from the photoionization models to the observations, we are left with a small set of solutions that agree on all three counts mentioned above. 


\section{Results}
\label{sec:results}

\subsection{First analysis}
\label{sec:first}


\begin{figure*}
\centering
\includegraphics[width=0.75\textwidth]{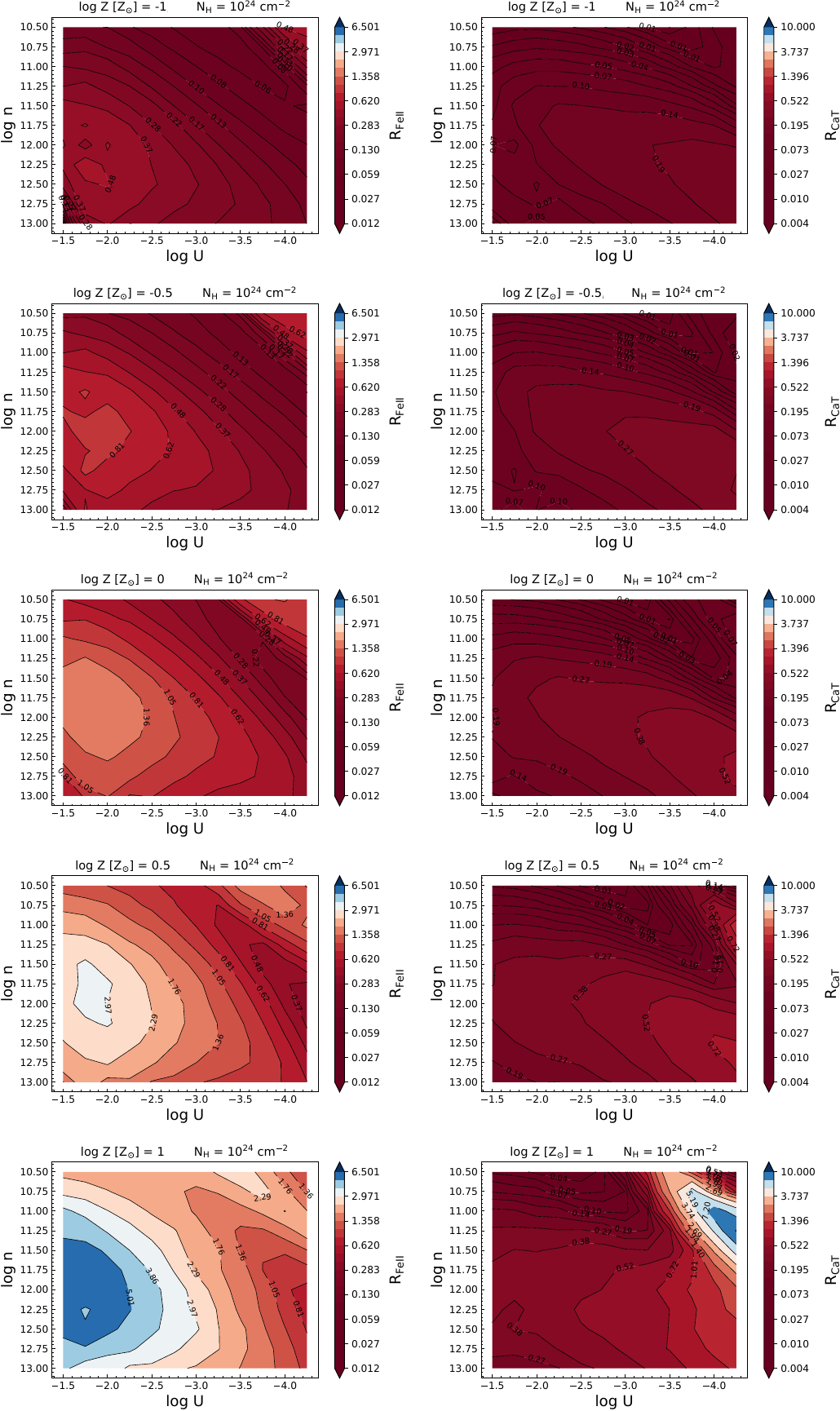}
\caption{LEFT: \un{} 2D histograms color-weighted by \rfe{} with column density, N$_{\rm{H}} = 10^{24}\,\rm{cm^{-2}}$ shown as a function of increasing metallicity (in log-scale and in units of \zsun{}). RIGHT: Physical parameters are identical to the left panels. Plots are color-weighted by \rcat{}.}
\label{fig:un_rfe2_rcat_24}
\end{figure*}


The results from our base setup is shown in Figure \ref{fig:un_rfe2_rcat_24}. The panels in this figure show the \un{} parameter space color-coded as a function of the flux ratios (\rfe{} or \rcat{}). The five panels show the setup as a function of increasing metallicity content considered in the BLR cloud model, i.e. Z = 0.1\zsun{}, 0.3\zsun{}, \zsun{}, 3\zsun{} and 10\zsun{}. These figures are constructed from models that utilize a cloud column density, N$\rm{_{H}}$ = $10^{24}$ cm$^{-2}$. We illustrate the effect of other cloud column densities in Sec. \ref{co-dep}. Note that \textit{U} represents the ionization parameter of the medium, and we later discuss whether its value should be estimated directly from the {\it observed} continuum.

In Figure \ref{fig:un_rfe2_rcat_24}, for the lowest metallicity case (log Z [\zsun{}] = -1, top-left panel), the maximum \rfe{} recovered is $\sim$ 0.575 (for log \u{} = -1.75, log \n{} = 12.25). For log Z [\zsun{}] = -0.5, this maximum rises to $\sim$ 0.906 (for log \u{} = -1.75, log \n{} = 12). This value of maximum \rfe{} further increases when the metallicity is raised to solar and super-solar values. At solar metallicity, the maximum \rfe{} recovered is $\sim$ 1.742 (for log \u{} = -1.75, log \n{} = 11.75), which is quite close to the estimate for \textmyfont{I Zw 1} by \citet{persson1988}, i.e. 1.778$\pm$0.050. To recover the \citet{murilo2016} \rfe{} estimate, i.e. 2.286$\pm$0.199, we assume metallicity values that are higher than solar values. For log Z [\zsun{}] = 0.5 and log Z [\zsun{}] = 1, we recover values for \rfe{} $\sim$ 3.296 (for log \u{} = -1.75, log \n{} = 12) and $\sim$ 6.501 (for log \u{} = -1.5, log \n{} = 12). Hence, from this base model analysis, we find that we can indeed recover the \rfe{} estimates that are consistent with the highest \feii{} emitters under the assumption that the average metal content in the BLR cloud be \zsun{} $\lesssim \rm{Z} \lesssim$ 3\zsun{}. 

Similarly, for the \rcat{} cases (right panels in Figure \ref{fig:un_rfe2_rcat_24}) - for the lowest metallicity case, log Z [\zsun{}] = -1, the maximum \rcat{} recovered is $\sim$ 0.264 (for log \u{} = -4.25, log \n{} = 12.5). This value of \rcat{} is obtained for the log \u{} that is at the grid boundary. In order to assess this issue, we ran a sub-grid, going down by 1 dex till log \u{} = -5.25. For values lower than log \u{} = -4.25, we start to see a saturation and the recovered \rcat{} begins to plummet after this boundary value of -4.25. Hence, we keep this limit as it is and proceed further with our analysis. For log Z [\zsun{}] = -0.5, this maximum rises to $\sim$ 0.389 (for log \u{} = -4.25, log \n{} = 12.75). This value of maximum \rcat{} further increases when the metallicity is raised to solar and super-solar values. At solar metallicity, the maximum \rcat{} recovered is $\sim$ 0.557 (for log \u{} = -4.25, log \n{} = 12.5), which is quite close to the estimate for \textmyfont{I Zw 1} reported by both \citet{persson1988} and \citet{murilo2016}, i.e. 0.513$\pm$0.130 and 0.564$\pm$0.080, respectively. Requesting higher than solar metallicities in the case of \rcat{} recovers values that are yet to be confirmed observationally. Hence, from this base model analysis, we find that we can indeed recover the \rcat{} estimates that are consistent with the observed estimates if we request the metallicity to be of the order of solar values.

\subsection{The problem of the EWs and covering factors in LIL region}

These results are in-line with our conclusions obtained in P20. It was shown in P20, that our photoionization models can predict \rfe{} and \rcat{} based on their flux ratios and the modelled estimates were found to be in-line with the measured values from an up-to-date observational sample of 58 sources (see Table 1 in P20), and the measured correlation (almost one-to-one) between the two ratios were matched by both the modelled and observed data. We re-affirmed this in the previous section with the agreement extended to the radial distance of the BLR in terms of the emitting regions of these two ions.

However, the proper model should reproduce not only the line ratios but also the line intensities, reflected in the line EWs. Therefore we now also measure the line EWs from the grid of models. To do so, we utilise continua possibly close to the lines in consideration. For estimating the EWs for \hb{} and \feii{} we use one of \textmyfont{CLOUDY}'s default continuum values, i.e. at $\lambda$ = 4885.36\AA. We checked for differences with the usually considered continuum level that is at 5100\AA~ and found good agreement (they differ by $\sim$0.2\%). On the other hand, for the \cat{} emission, the triplet is located in the NIR part of the spectrum, and thus, needs a different continuum level to estimate the EWs properly. This has been employed in previous observational works \citep[see e.g.][]{martinez-aldamaetal15,murilo2016} which has to do with the additional contamination of the disk continuum by the reprocessed torus contribution. To mitigate this issue, we utilise another default \textmyfont{CLOUDY} continuum at $\lambda$ = 8329.68\AA, that is closer to the triplet and overlaps with the continuum windows in the NIR used in prior studies.

To estimate the covering factors for \feii{} and \hb{} predicted by \textmyfont{CLOUDY}, we first derive an average EW estimate from observations of a large sample of quasars that have similar physical properties to \textmyfont{I Zw 1}. We consider a subset of the DR14 Quasar Catalogue \citep[][]{rakshitetal2019} wherein the selected subset contains quasars that have FWHM(\hb{}) $\leq$ 4000 \kms{}, which are also referred to as Population A sources. Population A sources can be understood as the class that includes local NLS1s as well as more massive high accretors which are mostly classified as radio-quiet \citep{ms14} and that have FWHM(\hb{}) $\leq$ 4,000 \kms{}. Previous studies have found that the Population A sources typically have Lorentzian-like \hb{} profile shape \citep{sul02,zamfiretal10} in contrast to Population B sources, the latter have broader FWHM(\hb{}) ($\geq$ 4,000 \kms{}), are pre-dominantly ``jetted'' sources \citep{padovani2017} and have been shown to have \hb{} profiles that are a better fit with Gaussian (for sources with still higher FWHMs, we observe disk-like double Gaussian profiles in Balmer lines). This subset from \citet{rakshitetal2019} contains 48,017 sources (about 9\% of the total catalogues sources). In addition to the FWHM limit that limits the sources within the Population A type \citep[][and references therein]{mar18}, we employ a quality cut on the estimated EWs from the catalogues by limiting the errors associated with the EW(\hb{}) measurements within 20\%. This reduces the sample to 28,252 sources. The estimated mean and standard deviation values (in \AA) for this subset - for \feii{} = (48.84, 52.42) and for \hb{} = (68.13, 46.90). We also estimated the mean and standard deviation for the EWs for the H$\alpha$ measurements in this subset which gave us (299.05, 141.67). Going by the arguments for the typical values for the Balmer decrement in AGN (here, H$\alpha$/\hb{}) $\approx$ 3 \citep[see Figure 3 in][]{2008MNRAS.383..581D} we recover an average EW for \hb{} to be $\approx$ 100 \AA\ given the EW(H$\alpha$). Thus, for simplicity, we assume a generic value of 40 \AA\ for EW(\feii{}) and 100 \AA\ for EW(\hb{}) in our study. These generic values are confirmed by the sample of 58 sources in P20 containing the observations from \citet{persson1988,martinez-aldamaetal15,martinez-aldamaetal15b,murilo2016,2020arXiv200401811M}.

To compare the observed EWs to the model predictions we require a certain covering fraction (or factor) that scales the modelled EWs. The covering factors associated with the line species are grossly over-predicted -- for \hb{}, the derived EWs from the models are quite low, requiring covering factors $\gtrsim$100\% for 547 out of the 660 models (these 660 models include all 5 metallicity cases) to be comparable to the observed values. This implies that most of the model predictions of line intensities are low, by up to two orders of magnitudes recovered from these models. In the next section, we propose a way to mitigate this issue and to recover EWs from the models with reasonable covering factors.

\subsection{A simple proposition}

We consider three cases of covering factor for the LIL region that have a typical EW(\feii{}) = 40\AA~ recovered for Population A type sources - at 30\%, 45\% and at a more liberal 60\%. These values for the covering fractions of the LIL region are representative and agree with the values from the traditional single-cloud and the \textit{locally optimized cloud} (LOC) models \citep{baldwin1995, 2001ApJ...553..695K}. The need for high covering fractions is substantiated to explain the strengths of the emission lines in the BLR and the lack of the Lyman continuum absorption suggesting a flattened distribution of the BLR and the distant observer seeing the source at relatively small viewing angles \citep[][see also Figure \ref{fig:cartoon} in the current paper]{g09}. Past studies have suggested that the covering factors of the BLR and the torus are similarly based on the following reasoning - (A) if the torus has a lower covering factor than the BLR we would see the BLR in absorption against the central continuum source in some objects near the type-2 viewing position. This is never seen. (B) On the other hand, if the BLR has a lower covering factor, some regions of the dusty torus will see direct radiation from the central source. This cannot be the case for much of the torus because it would then be unable to exist as close in as is seen \citep[][and references therein]{2007arXiv0711.1025G,g09}. We thus assume here that the covering fractions for the two entities (the LIL region and the torus) are similar and substantiate the assumed covering fractions from prior statistical studies on large quasar samples to recover the covering fractions for the torus. Previous observational studies have estimated the covering factors, e.g. by utilising the ratio of IR to UV/optical luminosity \citep{2013MNRAS.429.1494R} for luminous type 1 (or unobscured) quasars from large surveys in those wavebands. They estimate a mean value for the covering factor 0.39$^{+0.23}_{-0.14}$. On the other hand, \citet{gupta2016} consider the ratio of the mid-IR luminosity to the bolometric luminosity to estimate the covering factors for a sample of radio-loud and radio-quiet sources. For their radio-quiet sample, they estimate a median covering factor $\sim$ 0.29. In addition to these estimates, \citet{Moretal2009} estimated the covering factor for \textmyfont{I Zw 1} to be $\sim$ 63\%. This was achieved by fitting Spitzer/IRS (2-35 $\mu$m) spectrum for the source using a clumpy torus model supplemented to models for dusty narrow line region clouds and dust, where the latter was modelled using a black-body distribution.

\begin{figure*}[hbt!]
    \centering
    \includegraphics[width=\textwidth]{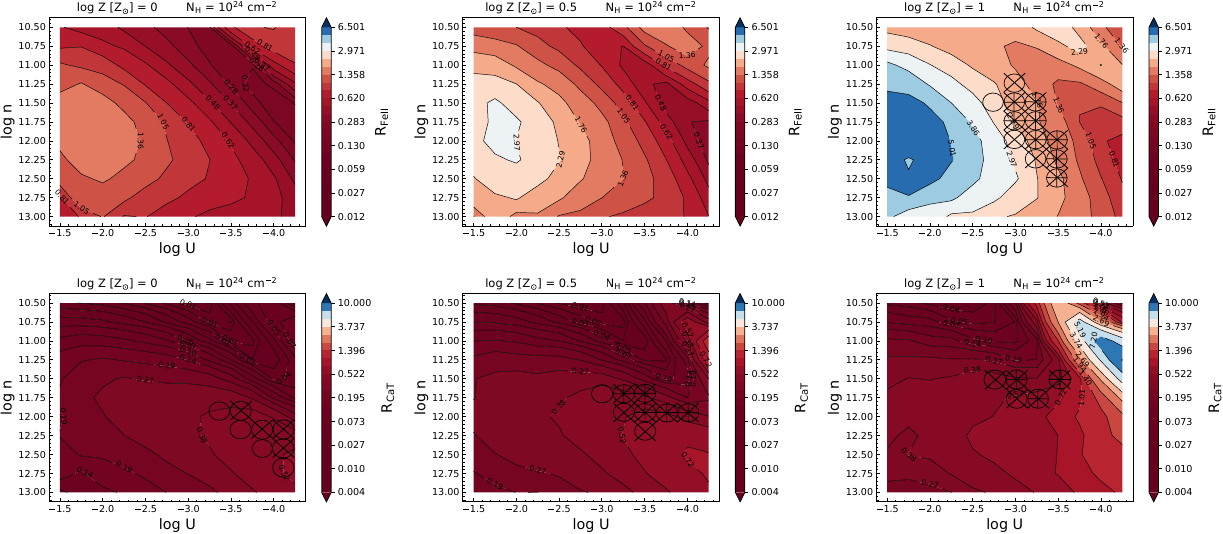}
    \caption{Three cases of metallicities (\zsun{}, 3\zsun{} and 10\zsun{}) models, previously shown in Figure \ref{fig:un_rfe2_rcat_24} for \rfe{} (upper panels) and \rcat{} (lower panels), respectively. Additionally, we overlay the solutions that are in agreement with corresponding lines' EWs considering three representative covering factors (30\% - `+', 45\% - `$\times$', and 60\% - `$\circ$').}
    \label{fig:ews}
\end{figure*}

\begin{figure}[hbt!]
    \centering
    \includegraphics[width=0.45\textwidth]{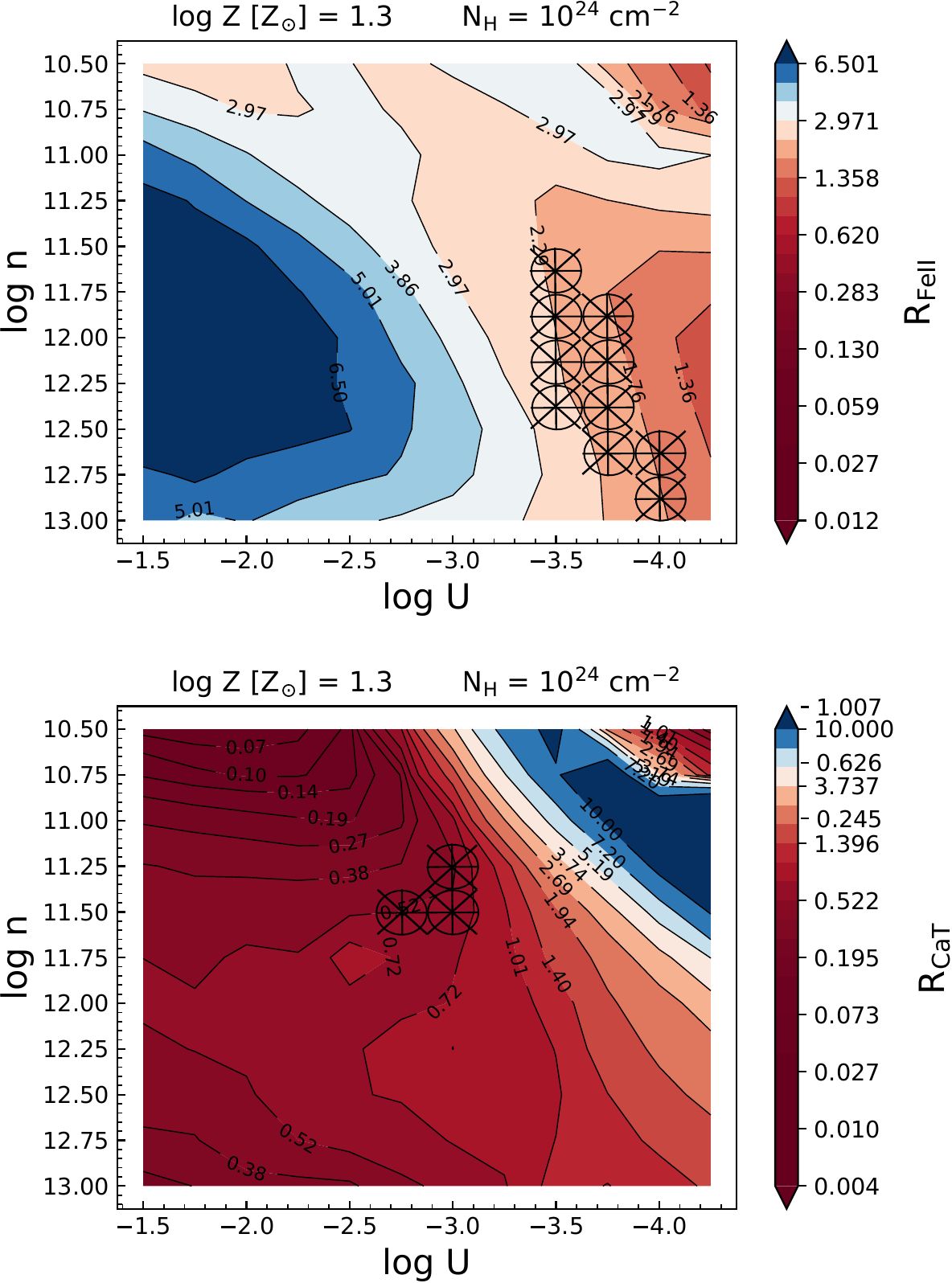}
    \caption{Same as Figure \ref{fig:ews} for Z $\approx$ 20\zsun{} for \rfe{} (upper panel) and \rcat{} (lower panel).}
    \label{fig:ews_z20}
\end{figure}

The location of these solutions that agree with the EWs in addition to the observed flux ratios are shown in Figure \ref{fig:ews} using special symbols\footnote{For lower covering factors (e.g. $\sim$10\%), we have one solution each for the \rfe{} and \rcat{}, i.e. at log \u{} = -3.5, log \n{} = 11.75 at log Z [\zsun{}] = 1 (for \rfe{}), and, log \u{} = -3.5, log \n{} = 11.5 at log Z [\zsun{}] = 1 (for \rcat{}). This pair of solution is unanimously recovered for all cases of the covering factors considered in this work.}. The underlying grid is identical to the respective panels shown already in Figure \ref{fig:un_rfe2_rcat_24}. We find that the solutions for \rfe{} are only plausible now at higher metallicities, of the order of $\sim$10\zsun{}. This is in-line with the observational evidences suggesting super-solar metallicities in excess of 10\zsun{} \citep[][]{hamannferland92,shinetal13,2020arXiv200914177S}. Models with lower metallicity values (Z$\lesssim$3\zsun{}) require covering factors that are above the requested limit ($>$60\%) and henceforth are not considered. On the other hand, for \cat{} emission, the flux ratios can still be produced from models that are at solar metallicities, although the covering factor required in such cases is higher ($\gtrsim$45\%). Increasing the metallicity to higher than solar, we have more optimal solutions in terms of low covering factor (see lower panels in Figure \ref{fig:ews}). For completeness, we also check for plausible solutions at higher than 10\zsun{}, by considering two additional cases - at 20\zsun{} and 100\zsun{}. We notice that in 20\zsun{} models (see Figure \ref{fig:ews_z20}), the solutions for \rfe{} are pushed to lower ionization parameters albeit at similar densities. There is limited solutions for the \rcat{} case that suggest not only radial sizes lower than \rfe{} by a factor 2, but smaller than the \hb{} reverberation mapping estimate. There are no solutions agreeing for any of the three chosen covering factors for the 100\zsun{} metallicity case. Hence, an increase in the metallicity up to $\sim$20\zsun{} values works well for \rfe{} estimates in the case of \textmyfont{I Zw 1}-like sources but not for corresponding \rcat{} emission. For the \rcat{} emission, metallicity values Z $\lesssim$ 10\zsun{} are found to be suitable to explain the EWs and the flux ratios. Summarising, the solutions that reproduce agreement on both optimal EWs and the flux ratios are obtained without much change in the density, log \n{} $\sim$ 11.75. However, the new solutions are now nearly two dexes lower in the ionization parameter for \rfe{}, i.e. log \u{} $\sim$ -3.5 (the maximum value for \rfe{} in the left panels of Figure \ref{fig:un_rfe2_rcat_24} correspond to a log \u{} $\sim$ 1.75).

Next, to assess the radial size of these emitting regions, we investigate the coupled distribution between the ionization parameter and local cloud density. As has been previously explored in \cite{negreteetal12, negrete2014, 2019Atoms...7...18M} and in P20, we take the product of the ionization parameter and the local cloud density (\u{}$\cdot$\n{}), i.e. this entity bears resemblance to ionizing flux, and for a given number of ionising photons emitted by the radiating source can be used to estimate the size of the BLR (\rblr{}): 

\begin{equation}
    R_{BLR}^{PM}\;[cm] = \sqrt{\frac{Q(H)}{4\pi Un_Hc}} \equiv \sqrt{\frac{L_{bol}}{4\pi h\nu\ Un_Hc}} \eqsim \frac{2.294\times 10^{22}}{\sqrt{Un_H}} 
    \label{eq1}
\end{equation}
where, \rblr{} is the distance of the emitting cloud from the ionizing source which has a mean local density, \n{}, and receives an ionizing flux that is quantified by the ionization parameter, \textit{U}. $Q(H)$ is the number of ionizing photons, which can be equivalently expressed in terms of the bolometric luminosity of the source per unit energy of a single photon, i.e. h$\nu$. Here, we consider the average photon energy, h$\nu$ = 1 Rydberg \citep{wandel99, marz15}. The specific value of the bolometric luminosity corresponds to 1 Zw I.


The \hb{}-based \rblr{} for \textmyfont{I Zw 1} was estimated to be $\sim$1.827$\times 10^{17}$ cm (= 37.2$^{+4.5}_{-4.9}$ light-days) obtained from the dedicated reverberation mapping campaign for this source \citep{huang2019}. The authors estimate the source to be a super-Eddington accretor with a dimensionless accretion rate \citep[$\mathcal{\dot{M}}$,][]{wang14} = 203.9$^{+61.0}_{-65.8}$. In order to validate the source's deviation from the standard \RL{} relation \citep{bentz13}, they estimate the contribution of the host galaxy and subsequently recover an AGN luminosity of $L_{5100}$ = 3.19$^{+0.27}_{-0.27}\times 10^{44}$ erg s$^{-1}$. This value for the source's AGN luminosity is well within the estimate from \citet{persson1988} within 1$\sigma$ uncertainty, i.e., $L_{5100}$ = 3.48$\times 10^{44}$ erg s$^{-1}$. The position of this source on the \RL{} diagram (see Figure 4 in \citealt{huang2019}) is almost at the boundary of the quoted scatter in the \RL{} relation by \citet{bentz13}, i.e., at 0.19$\pm$0.02 dex. This is also reflected in the conclusion of \citet{huang2019} that the source follows the empirical \RL{} relation.

On the other hand, the super-Eddington accretors have been found to show shorter lags compared to the their low-accreting counterparts which brings into question the validity of the standard \RL{} relation for these sources \citep{dupu2016L,2020MNRAS.491.5881Y}. Corrections have been proposed to the standard \RL{} relation, for example, linking to a dependence on accretion rate \citep{dupu2016L, martinez-aldama2019} and suggestion of a ``new'' \RL{} relation are being put forward that utilizes observables such as \rfe{} \citep{Du_Wang_2019} and \rcat{} (Mart\'inez Aldama et al., 2021, submitted). We test the hypothesis whether this new \RL{} relation including \rfe{} is suitable for \textmyfont{I Zw 1}. We utilize the two epochs of spectral information containing the \rfe{} estimate from \citet{persson1988} and \citet{murilo2016}, i.e. 1.778$\pm$0.050 and 2.286$\pm$0.199, respectively. For the earlier epoch, we set the $L_{5100}$ luminosity to 3.48$\times 10^{44}$ erg s$^{-1}$ and recover the \hb{}-based \rblr{} $\approx$18.68 light-days for \rfe{}$\approx$1.778. Next, for the more recent epoch, we set the $L_{5100}$ luminosity to 3.19$\times 10^{44}$ erg s$^{-1}$ as per \citet{huang2019} and get the \rblr{}$\approx$11.93 light-days for the \rfe{}$\approx$2.286. Both these \rblr{} estimates indicate that the lags thus obtained from this new \RL{} relation are shorter by upto a factor 3 than the lag value reported for \textmyfont{I Zw 1} by \citet{huang2019}. With this in mind, we consider that the standard \RL{} applies for this source and proceed accordingly.

Next, the similarity in the location of the emitting region for the \feii{} and \hb{} has been studied previously. The proximity of the ionization potential for the two ions suggests that they be produced in relatively close-by regions under similar physical conditions \citep[][and references therein]{panda18b}. \citet{hu15} found that the time delays of \hb{} and optical \feii{} are mostly similar, although there is scatter in their FWHM correlation that may suggest that the \feii{} is emitted from a larger region relative to the \hb{}. We, therefore consider the emitting region for the \feii{} and \hb{} to have significant overlap and for simplicity use the \hb{} radius obtained from \citet{huang2019} as a proxy for the \feii{}.


Now, with the agreement on the flux ratios for \feii{} and \cat{}, and their EWs in harmony with the observational evidence, we are left with the problem of matching the radial distances from the continuum source. We have a discrepancy between the radius of the emitting region suggested by the reverberation mapping and the one obtained by the photoionization method. In our case, the value of the radial distance obtained using Eq. \ref{eq1} for the physical parameters, log \u{}=-3.5, log \n{}=11.75, is 1.720$\times 10^{18}$ cm. We call this radius R$\rm{_{BLR}^{PM}}$. This value reproduces the EWs for the LILs as well as the lines' flux ratios in agreement with the observed values. On the other hand, from the \hb{} reverberation mapping of \textmyfont{I Zw 1}, we have the radial distance at 1.827$\times 10^{17}$ cm. We call this radius R$\rm{_{BLR}^{RM}}$. To match these radii and recover the expected ionization parameter, the luminosity incident on the BLR cloud responsible for the line reverberation needs to be scaled. We perform this scaling by employing the scaling relation between the radius-luminosity that agrees to the low-ionization emitting region, i.e. R$_{\rm{BLR}} \propto \rm{L^{0.5}}$. Hence, we have the following relation:

\begin{equation}
    L^{\prime} = L \times \left(\frac{R\rm{_{BLR}^{RM}}}{R\rm{_{BLR}^{PM}}}\right )^2
\end{equation}
where, \textit{L} and \textit{L$^{\prime}$} are the monochromatic luminosities for the photoionization and the reverberation method, respectively. Thus, in our case, this scaling value is $\sim$0.011 (i.e. only about 1\% of the original AGN continuum irradiates the BLR). This simply translates back to the lowering of the ionization parameter by nearly 2 dexes in our photoionization modelling that reproduce the flux ratios, the EWs within reasonable covering factors and scales down the \feii{} radial distance from the central source in agreement to the reverberation measurements. The implications for this filtering are discussed in Sec. \ref{sed}.

\subsection{Salient features of the \feii{} and \cat{} emission from photoionization}


\begin{figure*}[hbt!]
\centering
\includegraphics[width=\textwidth]{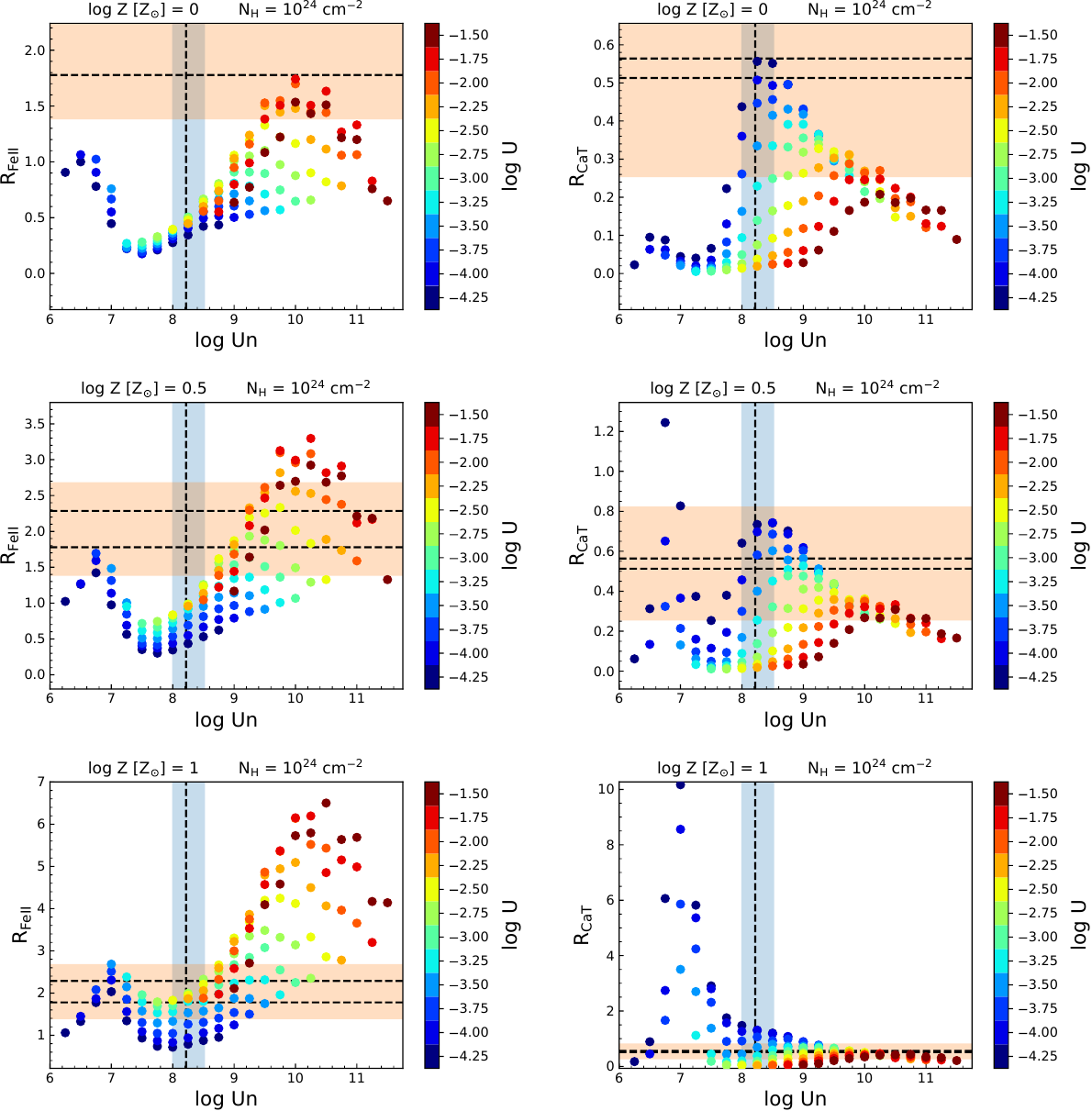}
\caption{Non-monotonic behaviour of \rfe{} versus log \u{}\n{} color-coded with respect to log \u{} (left panels). Corresponding cases for \rcat{} are shown in the  right panels. The panels represent the three sets of high-metallicity cases: log Z [\zsun{}]: 0 (top), 0.5 (middle) and 1 (bottom). The pale-blue vertical strip identifies the $Un_{H}$ product yielding the reverberation-based \rblr\ (black dashed line) within the uncertainties. The pale-orange horizontal strip identifies the observed \rfe{} and \rcat{} values (black dashed lines) from \citet{persson1988} and \citet{murilo2016} within uncertainties. Column density, N$_{\rm{H}} = 10^{24}$ cm$^{-2}$ is assumed.}
\label{fig:un_rfe2_24_zcases}
\end{figure*}

Since the consideration of the proper values of the line, EW has led us to an entirely different parameter space in the ionization parameters than in P20 and in Section~\ref{sec:first}, we now change our approach to search for the proper parameter space. We perform a three-step refining to extract the final solutions for the \un{} pertaining to the two parameters, \rfe{} and \rcat{}. Step 1 is matching the EWs for the \hb{}, \feii{} and \cat{} simultaneously within the requested covering factors (30\%, 45\% and 60\%). Then, the selected solutions are gauged against the radial distance that is within 20\%\footnote{The reported time delay for \textmyfont{I Zw 1} in \citet{huang2019} has an associated mean uncertainty of $\sim$13\%.} of the value obtained from the \RL{} for the \textmyfont{I Zw 1}'s luminosity ($\sim 3.48\times 10^{44}\;\rm{erg\;s^{-1}}$, \citealt{persson1988}). The last step of refining is matching the modelled estimates with the observed line flux ratios for both the ions, \rfe{} and \rcat{}. This is what gives us the solution marked atop the simulation grid in Figures \ref{fig:ews} and \ref{fig:ews_z20}.

The solutions which fully satisfy the observational constraints are a small sub-section of the original grid, as illustrated in Figures \ref{fig:ews} and \ref{fig:ews_z20}. To demonstrate better why P20 solutions with much higher ionization parameters were favoured we replot all solutions
    in Figure \ref{fig:un_rfe2_24_zcases}. The grid points from three panels for metallicity at \zsun{}, $\sim$3\zsun{} and 10\zsun{} for both \rfe{} and \rcat{} are extracted from the \un{} space and reported here in terms of the radial distance (as referred to in previous sections, the product of \u{} and \n{} for a fixed ionizing continuum gives the size of the line emitting region) versus the two flux ratios. The grid points are colour-coded with the corresponding ionization parameters. First considering the \rfe{} cases (left panels in Figure \ref{fig:un_rfe2_24_zcases}), we can see that the maximum emission in \rfe{} is nearly 2 dexes larger suggesting that the radial distance here is $\sim$10 times farther, which is what we explored in the previous sections. The vertical and horizontal patches in the plots indicate the \rfe{} estimates within 2$\sigma$ of the observed estimates and the radial sizes converted in \u{}\n{} scales. Here, $\sigma$ is taken as the maximum value of the error quoted from the two reported estimates from \citet{persson1988} and \citet{murilo2016}. Such a liberal range is considered keeping in mind that the observed and modelled estimates have subtle differences, such as, in the photoionization modelling with \textmyfont{CLOUDY}, the code considers 371-level accounting for $\sim$68,635 transitions for the \feii{} atom and are evaluated only up to $\sim$11.6 eV \citep{verner99}. In the analysis of the optical spectrum for \textmyfont{I Zw 1}, there is a need to supplement the fitting procedure with additional broad Gaussians in addition to the \feii{} pseudo-continuum generated from \textmyfont{CLOUDY} to minimize the residuals \citep[][Panda and Mart\'inez-Aldama in prep.]{negreteetal12}. This difference is highlighted in the subtle differences in certain \feii\ line transitions belonging to the \textit{$^4$F} group (mostly the 37 and 38 multiplets in the 4550 \AA\ and 4580 \AA\ wavebands). \citet{kovacevic2010} mitigate this problem by supplying line intensities found in the observational spectrum of \textmyfont{I Zw 1} in addition to the \feii{} line transitions expected from standard photoionization involving line recombination and collisional excitation processes. We apply the same approach (values with 2$\sigma$ uncertainties) while evaluating the \rcat{} panels. The overlapping region between the vertical patch and the horizontal patch marks the acceptable region for the solutions to the \rfe{}. As it can be noticed from the three left panels, the solutions are in best agreement when the BLR cloud has metallicity Z=10\zsun{}. The gradual increase in overall modelled distribution with an increase in metallicity suggests that the BLR clouds indeed require an overabundance of iron. On the other hand, for the \rcat{} case, solutions with quite low ionization parameters can successfully achieve the required \rcat{} estimates. Unlike the  \rfe{} cases, here \rcat{} can be modelled with a wider range of metallicities, \zsun{}$\lesssim$Z $\lesssim$10\zsun{}. Although in the higher-than-solar metallicity cases, the solutions that belong to the inter-junction of the appropriate radial distances and observed \rcat{} values in the plots show an increasing number of solutions that prefer higher ionization parameters (log \u{} $\gtrsim$ -4.0). These trends reveal that the two species (\feii{} and \cat{}) have a significant overlap in their emitting regions, although the results from this analysis suggest that the BLR cloud needs to be selectively overabundant in iron for optimizing the \feii{} emission. In contrast, sufficient \cat{} emission can be produced in a wider range of abundances ranging from solar to super-solar values. This points toward different formation channels for the two species, since iron is predominantly formed out of Type Ia SNe with CO-rich white dwarf progenitors while calcium which is an $\alpha$-element is mostly produced by Type II SNe after the explosion of massive stars \citep[7 \msun{} $\lesssim M_{*} \lesssim$ 100 \msun{},][]{hamann-ferland1993}. This aspect of the study is explored in detail in a separate work \citep{2021arXiv210106999M} using our observational sample compiled in P20.

\subsection{Co-dependence of metallicity and cloud column density}
\label{co-dep}

\begin{figure*}[hbt!]
\centering
\includegraphics[width=0.45\textwidth]{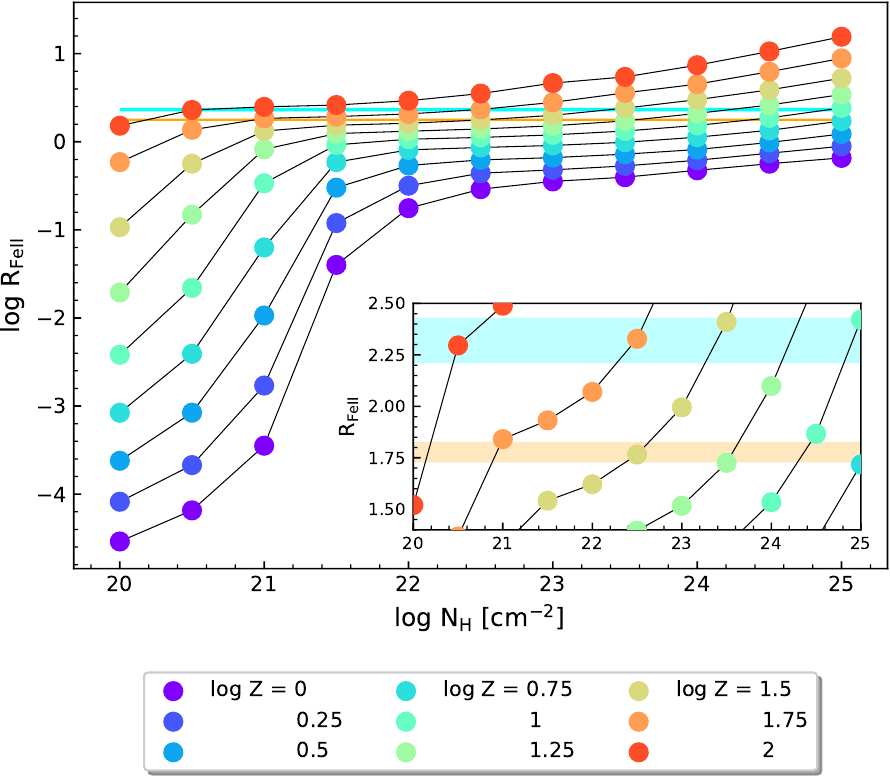}
\quad
\includegraphics[width=0.45\textwidth]{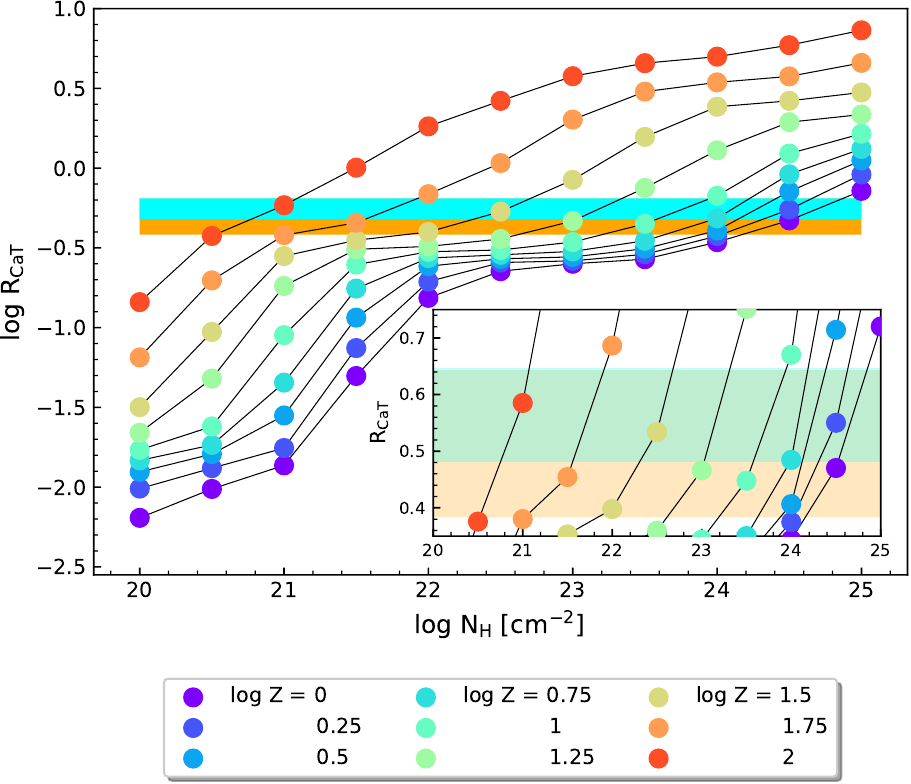}
\caption{LEFT: Strength of the optical \feii{} emission (\rfe{}) shown with respect to the distribution in cloud column density (N$_{\rm{H}}$) from \textmyfont{CLOUDY}. The model uses a log \textit{U} = -3.5 and log n = 11.75. The colors represent 9 different cases of metallicity (Z). The observed estimates from \citet{persson1988} and \citet{murilo2016} for \textmyfont{I Zw 1} are shown in cyan and orange bars (the errors in these estimates are depicted by the bar-width), respectively. The inset plot zooms in on a portion of the base plot to highlight the modelled trends that recover the \rfe{} within the observed values. Notice that the \rfe{} is shown in log-scale in the base plot while for the inset plot we have shown the ratio in linear-scale. RIGHT: Corresponding \rcat{} distribution for the same modelled parameters as in the previous panel.}
\label{fig:Z-N}
\end{figure*}

In P20, we explored, in a rather limited manner, the increasing trend of \rfe{} and \rcat{} estimates as a function of increasing column densities. We considered two additional cases in column densities apart from the base value of N$\rm{_{H}} = 10^{24}$ cm$^{-2}$, i.e., at $10^{24.5}$  and $10^{25}$ cm$^{-2}$, limiting our models within the realms of the optically thin regime, i.e., the optical depth ($\tau = \sigma_{\rm{T}}\cdot \rm{N_{H}}$), $\tau \sim 1-2$ for optically thin medium, which implies $\rm{N_H} \sim 10^{24} - 10^{24.5}\;\rm{cm^{-2}}$. Here, $\sigma_{\rm{T}}$ is the Thompson's scattering cross-section and $\rm{N_{H}}$ is the cloud column density. There is a clear hint that the real scenario perhaps points towards a collective increase in both metallicity and cloud column density. This supports the argument towards the use of very high metallicities (Z $\gtrsim$ 5\zsun{}) to recover the \rfe{} estimates for the strong \feii{} emitters \cite[][]{nagao2006,negreteetal12,2020arXiv200914177S} which has strong implications for the BLR cloud properties, especially their density and radial distributions. In this section, we test this connection between the two aforementioned parameters in terms of the \rfe{} and \rcat{} estimates they recover.

From the analyses in the previous sections, the pair of ionization parameter and local cloud density, i.e. log \u{}\n{}, that best reproduce the \rfe{} and \rcat{} in agreement to the observed flux ratios, keeping the BLR cloud within the limits of the \rblr{} as estimated from the reverberation mapping and constrained for the EWs (even at covering factor $\sim$10\%), is $\sim$-3.5 and $\sim$11.75 (cm$^{-3}$), respectively. We, therefore, fix these two values in the subsequent modelling and study the effect of the metallicity within \zsun{} $\leq$ Z $\leq$ 100\zsun{} with a step size of 0.25 dex (in log-space) and the cloud column density within $10^{20} \leq$ N$\rm{_{H}} \leq 10^{25}$ with a step size of 0.5 dex (in log-space). Here, the modelled range for metallicity is extended to higher values than assumed before to test their relevance in the BLR LILs emission. As for the cloud column density, N$\rm{_{H}} = 10^{23} - 10^{24}$ cm$^{-2}$ is often the norm to account for the adequate emission from the LILs where the situation is relatively less dynamic compared to the low-density High Ionization Lines (HILs). The HILs require a higher ionization parameter and hence are proposed to originate much closer to the black hole and bear a more direct continuum as opposed to the LILs \citep{leighly04,negreteetal12, martinez-aldamaetal15}. Also, at the expected radial extensions for the LILs, the cloud is relatively cold to clump together and the lowering of the net radiation pressure keeps the cloud relatively extended \citep{2008ApJ...678..693M, 2009ApJ...695..793N}. On the other hand, having a larger cloud column allows for species like \feii{} to increase their ionic fraction to \hb{} and thereby produce enough emission to account for the \rfe{} $\gtrsim$ 1 as often seen for the high \feii{} emitters belonging to the extreme Population A \citep[see][and references therein]{bv08,panda18b,panda19b}.

In Figure \ref{fig:Z-N}, we demonstrate the strong degeneracy between the two quantities (metallicity and cloud column density) as a function of the recovered values for \rfe{} and \rcat{}. It is easily noticed that the same value of \rfe{} or \rcat{} can be derived by widely different combinations of the column density and metallicity. The main plots are in log-log space to appreciate the large extent of the intensity ratio against the 5 order stretch of cloud column density. From prior spectroscopic observations for \textmyfont{I Zw 1}, the \rfe{} and \rcat{} estimates have been reported -- (a) \rfe{} and \rcat{} estimates from \cite{persson1988}: 1.778$\pm$0.050 and 0.513$\pm$0.130, respectively; (b) \rfe{} and \rcat{} estimates from \cite{murilo2016}: 2.320$\pm$0.110 and 0.564$\pm$0.083, respectively. We utilize these measurements and overlay them on Figure \ref{fig:Z-N} with the quoted uncertainties in the measured values. For \rfe{} case, the models that have metallicities Z $\lesssim$ 3\zsun{} can't account for the expected intensity ratio, not even for the lower limit from \citet{persson1988}, even at the highest column density ($10^{25}$ cm$^{-2}$) considered in the analysis. We start to enter the optimal regime with Z $\sim$ 5\zsun{} and higher. The inset plot zooms in on the optimal range of solutions in terms of the \rfe{} recovered (note the linear scale used here for the y-axis), and, the needed column density and metallicity value to obtain that value. In principle, BLR cloud with the smallest size can reproduce the optimal \rfe{} emission, although in this case, the models require extremely high metallicity (100\zsun{}). Such an inverse behaviour between the metallicity and cloud column size isn't a surprise since these clouds are effectively made of mostly hydrogen and helium that exist in the front-facing part (or the fully ionized zone) of the cloud, while heavier and more metallic elements tend to occur in deeper parts of the cloud as revealed by the increase in the ionic fractions for the latter as a function of the depth within the BLR cloud \citep[see Figure 4 in][]{negreteetal12}. As we increase the column size, the \rfe{} estimate can still be obtained with lower metallicity values. For \rcat{}, the trend between the \rcat{} and cloud column density is rather monotonic in log-log space. Similar to the \rfe{}, smaller cloud sizes suggest higher metallicity, yet solutions with almost solar values for metallicity are sufficient to recover the required \rcat{} emission for cloud column sizes that are similar to the \rfe{} case, i.e. N$\rm{_{H}} \gtrsim 10^{24}$ cm$^{-2}$. Hence, a degeneracy between these two quantities, metallicity and cloud column density sustains. We discuss this conundrum between the metallicity and cloud column density in the BLR clouds and highlight ways to break this degeneracy in Section \ref{sec:discussion}.

\subsection{Microturbulence: a metallicity controller?}

\begin{figure*}[hbt!]
\centering
\includegraphics[width=\textwidth]{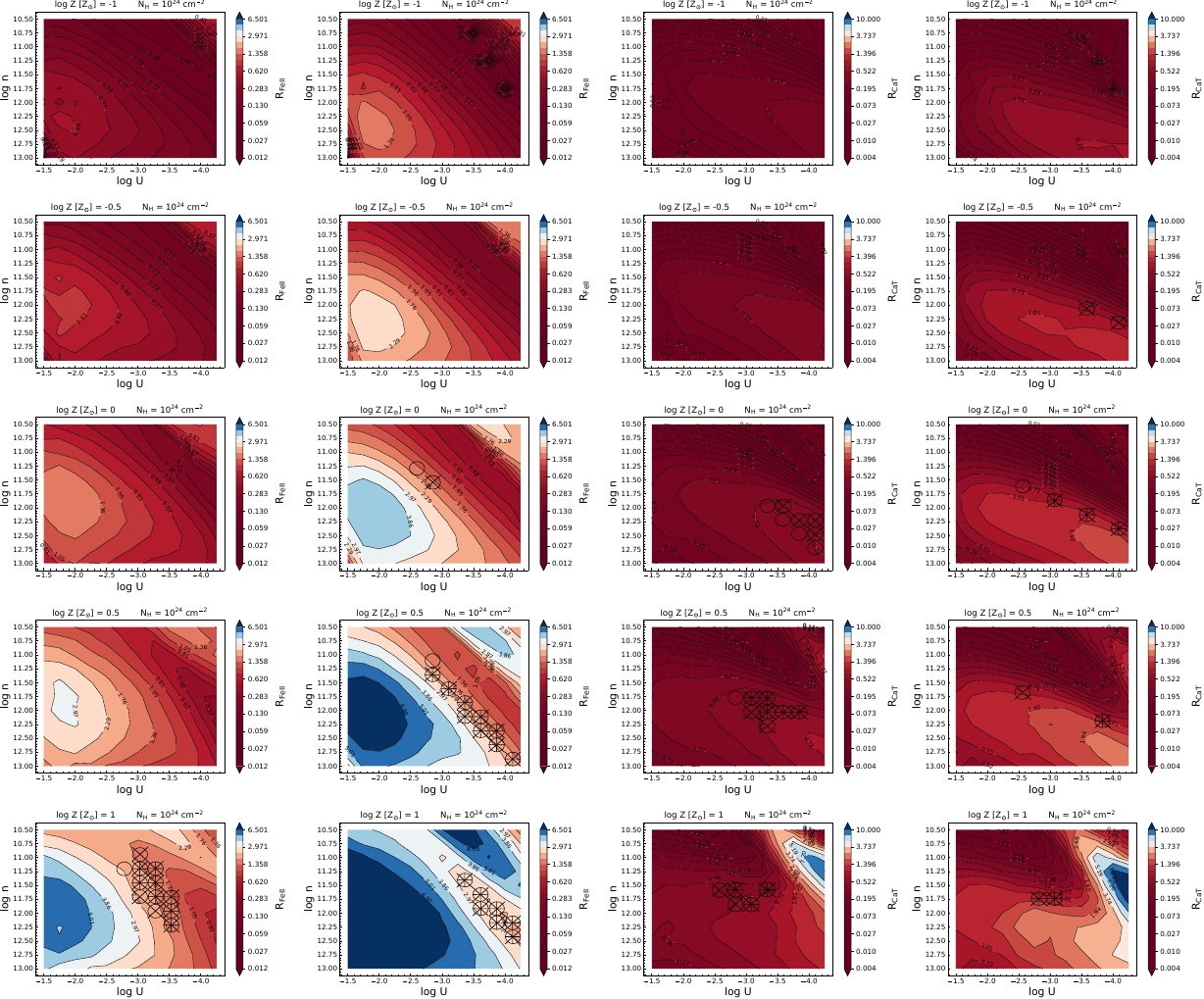}
\caption{Effect of microturbulence: The first and the third columns are from the original models without any microturbulence for \rfe{} and \rcat{}, respectively. The second and the fourth columns are the corresponding cases with microturbulence = 20 \kms{}. Each column consists of the five cases of metallicities considered in this work. The overlaid symbols have the same meaning as shown previously in Figure \ref{fig:ews}.}
\label{turb20}
\end{figure*}

Another important aspect to optimize the \feii{} emission is the effect of the microturbulence that has been noted to provide additional excitation \citep{baldwin2004,bv08}. The velocity field around a black hole might be a superposition of different kinematic components, such as Doppler motions, turbulence, shock components, in/outflow components, and rotation. Different velocity components result in different profiles, and the final profile is a convolution of different components \citep{kol13}. And, local turbulence substantially affects the \feii{} spectrum in photoionization models by facilitating continuum and line–line fluorescence. Increasing the microturbulence can increase the \feii{} strength and give better agreement between the predicted shape of the \feii{} blends and observation \citep{shi10}. The effect of the microturbulence has been carefully investigated in our previous works \citep{panda18b,panda19}, where a systematic rise in the \rfe{} estimates, is obtained by increasing the microturbulence up to 10-20 \kms{}. After this limit the \rfe{} tends to drop and for 100 \kms{} this reaches values similar to zero microturbulent velocity. We test the effect of the microturbulence in the context of this study, especially if this entity works similarly for boosting the \cat{}. We consider a microturbulence value of 20 \kms{} and re-run our models. The results are summarized in Figure \ref{turb20} for the two ions side-by-side. As expected, for the \rfe{} case, the microturbulence effect yield comparable \rfe{} estimates for lower metallicity. For example, the case with no microturbulence with solar metallicity gives \rfe{} values similar to the microturbulence = 20 \kms{} at 0.3\zsun{}. This effect is seen in other metallicity cases as well. For the preferred solution with $\sim$10\zsun{} for the case with no microturbulence, upon invoking this parameter, we achieve the solution with the $\sim$3\zsun{} models. On the other hand, for the \rcat{} cases, the results are almost similar between the two versions, indicating that the \cat{} emission is probably less prone to fluorescence effects. We overlay the solutions that agree with the lines EWs for the three cases of covering factors similar to Figures \ref{fig:ews} and \ref{fig:ews_z20}. For a much lower covering factor ($\sim$10\%), we find that with the inclusion of microturbulence in the medium, \rfe{} estimates closer to \citet{murilo2016} are more probable with ionization parameters log \u{}$\sim$-3.5, and densities log \n{}$\sim$11.5, albeit at 10\zsun{}. For the same low covering factor, there is a unique solution satisfying for \rcat{}, i.e. for log \u{}$\sim$-3.25 and log \n{}$\sim$11.5 also at 10\zsun{}, which shows that the two ions can have significant overlap in their emitting regions. This is another confirmation of the nearly 1:1 correlation obtained in P20 between the \rcat{} and \rfe{}. There are solutions with higher covering factors that agree with the line equivalent widths at metallicities Z$\lesssim$10\zsun{} in the \rfe{} cases with ionization parameter as high as log \u{}$\gtrsim$-2.75, yet the cloud densities remain nearly unchanged at log \n{}$\gtrsim$11.5 (cm$^{-3}$). These latter solutions then require larger covering factors ($>$30\%) to account for the optimal emission in \feii{}. The last panel of \rfe{} cases with microturbulence included (10\zsun{}) has a significant overlap with the solutions realised from Z$\approx$20\zsun{} models for \rfe{} (see Figure \ref{fig:ews_z20}). The effect of microturbulence is only a secondary effect seen from the spectra as this affects mostly the wings of the broad line profiles \citep{goad12}, one that becomes quite difficult to estimate properly as these features become increasingly close to the noise level.

\section{Discussions}
\label{sec:discussion}


\begin{figure}[hbt!]
    \centering
    \includegraphics[width=0.825\columnwidth]{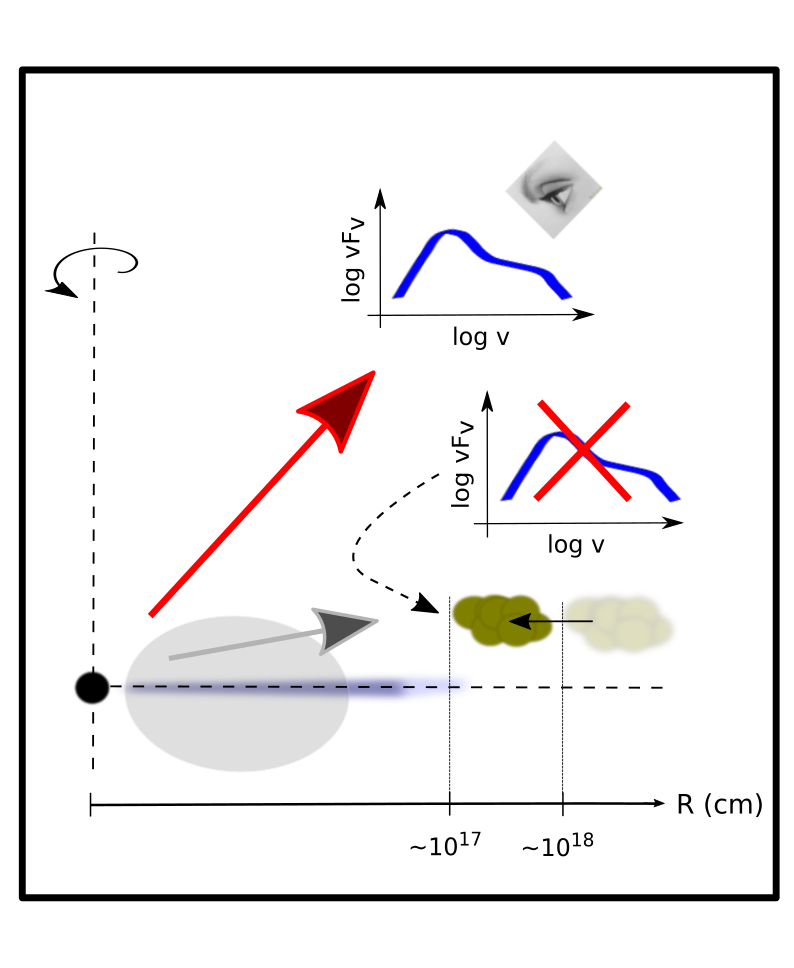}
    \caption{Schematic view of the anisotropy in the radiation between the observer and the BLR cloud. Our model considers a simple shrinking of the radial position of the BLR to match the reverberation-mapped \rblr{} estimate, by filtering the incoming radiation from the accretion disk. Thus the net SED seen by the BLR differs from the SED seen by a distant observer. This illustration applies to \textmyfont{I Zw 1}-like sources, i.e. Type-1 Narrow-line Seyfert galaxies with high \feii{} emission, which are the context of this study.}
    \label{fig:cartoon}
\end{figure}

The problem to reproduce the EW of LILs has been discussed in the literature before. Either additional mechanical heating is necessary \citep[e.g.][]{collin-souffrin1986,jol87}, or multiple cloud approach, with part of the radiation scattered/re-emitted between different clouds, or the BLR does not see the same continuum as the observer \citep{kor97} due to an intervening medium such as a wind component \citep{leighly04}. This wind component is often seen in high-ionization lines in the UV, such as \civ{}, of an AGN spectrum, typically belonging to the Population A type \citep[][and references therein]{mar18} like \textmyfont{I Zw 1}. In our computations, we recover the proper Fe II ratio {\it and} EW for low values of the ionization parameter which indicates that we do not need extra energy to produce stronger lines but we have to reduce the incident flux - otherwise the medium is over-ionized and line production efficiency drops. This favours the option of the intervening medium. We have applied the latter hypothesis to analyze \textmyfont{I Zw 1} and its LILs of the BLR. This scenario is also supported by recent observational findings in \citet{2020MNRAS.492.3580W}, wherein the authors have found that the \feii{}-emitting region is shielded from the central source for a sample of $\sim$2100 Type-1 AGNs \citep[see][for a similar inference on \cat{}]{martinez-aldamaetal15}. We illustrate this scenario in Figure \ref{fig:cartoon} wherein the key assumption is that the broad-band spectral energy distribution seen by the BLR is different from the one that is perceived by a distant observer. This hypothesis can be perceived as the combined effect from (a) geometrical effects, and (b) radiation filtering due to obscuration. Although other effects, such as lensing and limb darkening which modify the inclination dependence, could additionally supplement in recovering this radiation filtering. The flattened disk geometry results in an inclination dependent flux ($\propto$ cos$\theta$, where $\theta$ is the angle between the symmetry axis and the line-of-sight of the observer). On the other hand, the filtering, or rather the collimation of the continuum can be a result of the anisotropy in the disk's radial structure. The anisotropy originating from the accretion disk has been suggested by previous studies \citep[][and references therein]{wang14} pointing away from a generally assumed geometrically thin accretion disk, especially in the regions at close vicinity of the black hole. Such a geometry inhibits the radiation coming from the inner, hotter region, making it possible for the BLR to receives a continuum that is a fraction of the original ionizing flux. This is a valid assumption in the case where the observer is systematically at an offset in the viewing angle to the BLR cloud. 

\subsection{Analyzing the change in the shape of the SED}
\label{sed}

A primary result of the paper is that BLR clouds that have sizes of the order of $\approx10^{12}$ cm, need to see about 1\% of the SED and have optimized properties to replicate all the observational constraints in extreme objects like \textmyfont{I Zw 1}. Here, we discuss the implication and justification based on prior studies.

\begin{figure*}[hbt!]
    \centering
    \includegraphics[width=\textwidth]{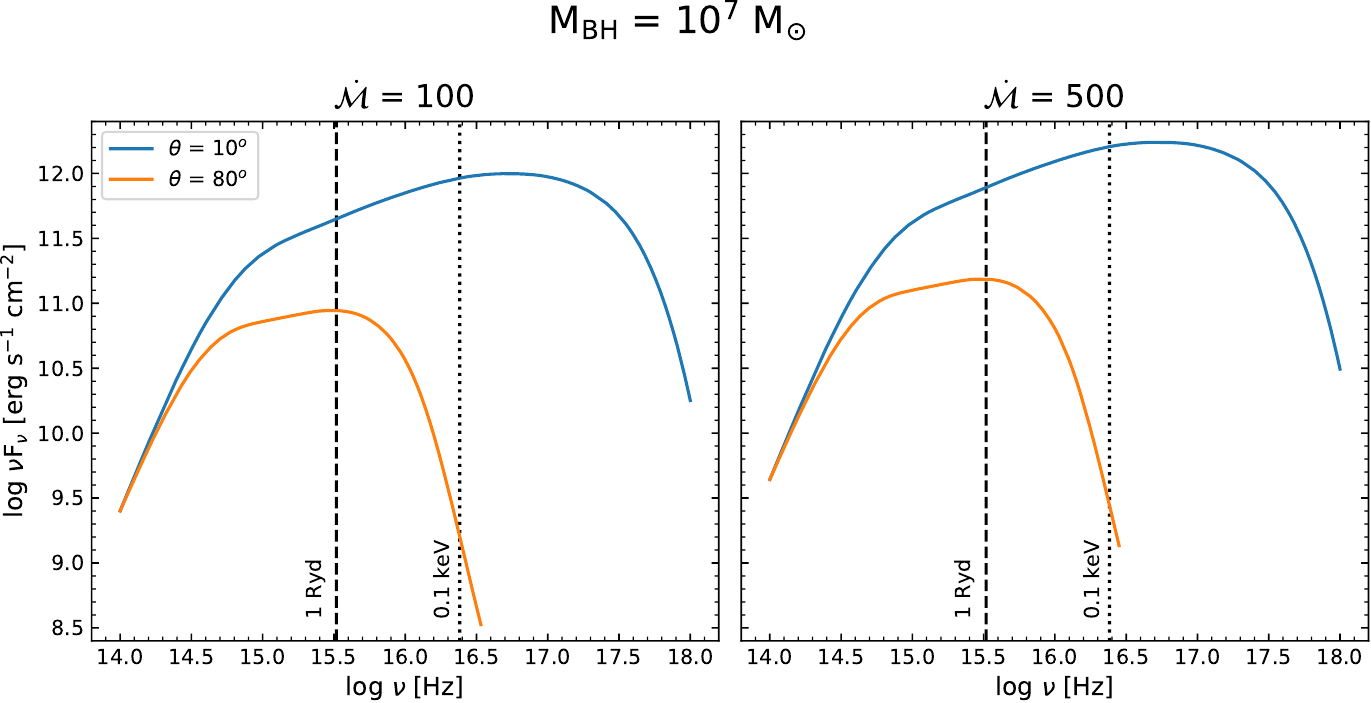}
    \caption{Spectral energy distributions obtained for slim accretion disks for a representative black hole mass, \mbh{} = 10$^7$ \msun{} for two cases of accretion rates (dimensionless): LEFT: \mdot{} = 100, RIGHT: \mdot{} = 500. In both panels, SEDs are shown for two cases of orientation, one representative for the distant observer, i.e. $\theta$=10$^{o}$ (shown in orange) and the other representative for the BLR, i.e. 80$^{o}$ (shown in blue), where $\theta$ is the angle between the line-of-sight and the black hole symmetry axis. The locations for the 1 Rydberg and 0.1 keV are highlighted in each panel with the black dashed and dotted lines, respectively. These models are taken from \citet{wang14} and are applicable for \textmyfont{I Zw 1}-like sources, where in particular, the $\mathcal{\dot{M}}$ for \textmyfont{I Zw 1} estimated by \citet{huang2019} is 203.9$^{61.0}_{-65.8}$.}
    \label{fig:compare_sed}
\end{figure*}

\citet{wang14} studies in detail the solutions for the structure of accretion disks from sub-Eddington accretion rates to extremely high, super-Eddington rates (where the dimensionless accretion rate, \mdot{} $>>$ 1). The appearance of sharper funnels in the innermost region (below 3$R_{g}$, where $R_{g}$ is the gravitational radius) as the accretion rate increases significantly modifies the thin-disk geometry that applies under the \citet{ss73} regime. The authors note that the aspect ratio (\textit{h} = \textit{H/r}, where \textit{H} is the height of the disk and \textit{r} is the distance along the radial direction both in the units of $R_{g}$ such that \textit{h} is dimensionless) of the funnel for these slim disks is insensitive to the black hole mass and the shape of a slim disk has three notable features: (1) a funnel that develops in the innermost region [dh/dr > 0]; (2) a flattened part [dh/dr < 0 and h$\sim$1]; and (3) a geometrically thin part
(h$\sim$10$^{-2}$), approaching the \citet{ss73} regime, in which the funnel disappears. They further investigate the effect of the SEDs received by the BLR (or a distant observer) with different inclinations to the disk. Here, the BLR clouds are considered to be quite close to the system's mid-plane (see Figure \ref{fig:cartoon}) such that the inclination angles (to the symmetry axis) subtended by these clouds are quite high, around 75-85$^{o}$\footnote{here the angle is estimated using the relation: 90$^{o}$ - tan$^{-1}\left(\frac{H'}{r'}\right)$, where \textit{H'} is the peak height attained by the BLR cloud and \textit{r'} is the corresponding radial distance of the BLR cloud from the black hole. This picture of the BLR considers clouds accelerated under the combined influence of radiation pressure and gravity. We refer the readers to \citet{2021arXiv210200336N} for more details.}. On the other hand, the face-on view of NLS1 sources such that the inclination angle subtended to the distant observer is relatively small \citep{2001ApJ...561L..59W,panda19b} is supported by the small widths of the Balmer lines due to the projection effect \citep{1985ApJ...297..166O,2004MNRAS.352..823B}. Recent studies by \citet{2017ApJS..229...39R} who used the SDSS DR12 data to construct an NLS1 catalogue also found that, statistically, the NLS1s have smaller viewing angles in comparison to their broad-line counterparts. In \citet{wang14}, the authors notice the anisotropy of the radiation field clearly with their study and note that:  (1) the flux received by the clouds (or the distant observer) dramatically
decreases with increasing inclination by a factor of 30 (i.e. going from $\theta$= 10$^{o}$ to $\theta$= 80$^{o}$), which is much steeper than what is recovered with the simple cos$\theta$ dependence; and (2) the SEDs are significantly softened by self-shadowing at higher inclinations, resulting in the lack of photoionizing photons required for the emission lines. We highlight the relative change in the bolometric flux content as a function of the source's inclination with respect to the distant observer in Figure \ref{fig:compare_sed} for two sets of slim disk SEDs (at \mdot{}=100 and 500, Jian-Min Wang, priv. comm.) at black hole mass = 10$^{7}$ \msun{} similar to the estimate from \cite{huang2019}, i.e. 9.30$^{+1.26}_{-1.38}\times 10^6$ \msun{}. The selection of the lower value of inclination was based on \citet{Moretal2009} estimate for the viewing angle from torus fitting to Spitzer/IRS 2-35 $\mu$m spectra for \textmyfont{I Zw 1} suggesting a value for $\theta \approx 8^{o}$. We chose a SED model with $\theta$=10$^{o}$ to mimic the SED seen by a distant observer. On the other hand, for the BLR, we choose a modelled SED with a higher inclination angle ($\theta$=80$^{o}$). We estimate the flux ratio between the two cases of inclination at 0.1 keV according to \citet{wang14} that drives the \hb{} emission. For the \mdot{} = 100 case, we obtained the flux ratio $\approx$ 96.67\%. For a higher accretion rate, e.g. \mdot{} = 500, we obtain a value ($\approx$ 99.92\%) consistent with the filtering factor we obtain from the scaling of the BLR radius. These flux ratios, when estimated at 1 Rydberg give us 62.5\% and 90.62\% for \mdot{} = 100 and 500, respectively.

The anisotropic emission from the accretion disk has been tested more rigorously in past studies \citep[][and references therein]{runnoe2013,xu2015,2016A&A...587A..13L}, such as in the context of estimating quasar bolometric corrections considering thin accretion disks as the source of the continuum emission including relativistic effects \citep{nemmen2010}. In such cases, the authors find a slump in the net integrated luminosity for a source over an order of magnitude when the viewing angle increases from $\theta$=10$^{o}$ to $\theta$=80$^{o}$ (see Figure 3 in \citealt{nemmen2010}). The relative decrease in this luminosity as a function of the increasing viewing angle can be even higher under Newtonian approximation where the bolometric luminosity is related to the integrated luminosity assuming isotropic emission such that: 

\begin{equation}
L_{bol} =  \frac{1}{2cos\theta}\int_{\nu_0}^{\nu_1}L_{\nu}\; d\nu
\end{equation}
where, $\nu_0$ and $\nu_1$ are the frequencies bounds for thin disk radiation. On the other hand, strong light bending effects in quasar microlensing events can cause differential lensing distortion of the X-ray and the optical emission and change significantly change the X-ray-to-optical flux ratios in such sources \citep{chen2013}.

Narrow line Seyfert galaxies with high accretion rates are typically shown to have a soft X-ray excess \citep{arnaud1985} in their broadband SED \citep{jin12a,jin12b,mar18,ferlandetal2020}. The interstellar medium blocks our view of this spectral region thus requiring the use of indirect methods to predict the emission from this part of the radiation field \citep[][and references therein]{kub18}. This component helps to bridge the absorption gap between the UV downturn and the soft X-ray upturn \citep{elvis94,laor97,richards06} and changes the far-UV and soft X-ray part of the spectrum thus affecting the \feii{} line production \citep{panda19}. Thus, there is a need to expand the study to construct more realistic SEDs for \textmyfont{I Zw 1} and similar sources which will be undertaken in a subsequent study.

\subsection{Degeneracies with the metallicity and cloud sizes in the BLR}
Additional constraints from high signal-to-noise rest-frame UV spectrum for \textmyfont{I Zw 1} can help narrow down the possibilities for the metallicity. There are quite a few metallicity indicators, for example, \aliii$\lambda$1860/\heii$\lambda$1640, a fairly unbiased estimator for the metallicity \citep[see][for an overview]{2020arXiv200914177S}. Another line ratio frequently used is \siiv$\lambda$1397+\oiv$\lambda$1402/\civ$\lambda$1549\ \citep[][and references therein]{hamannferland99}. The choice of diagnostic ratios used for metallicity estimates is usually a compromise between S/N, easiness of deblending, and straightforwardness of physical interpretation. \cite{laor971zw1} made the spectral decomposition of \textmyfont{I Zw 1}'s HST-FOS spectrum and reported the various spectral parameters in their paper. The \aliii/\heii\ flux ratio from their analysis is $\approx$ 1.78 and the \siiv+\oiv/\civ\ gives $\approx$ 0.89, suggesting a metallicity $\sim$10\zsun{} and slightly above solar, respectively. However, another ratio, \nv$\lambda$1240/\heii\ flux ratio gives a value $\sim$5.78 suggesting Z$\gtrsim$10\zsun{}, although this ratio is quite sensitive to change in ionization parameter \citep{2012ApJ...751L..23W}. Other ratios, such as \civ/\heii\ and \siiv+\oiv/\heii\, also point towards similarly high metallicities (Z $\gtrsim$ 10\zsun{}), although they are not so reliable due to issues related to blending with other species which becomes cumbersome unless a better quality spectra are available. Hence, utilizing the \aliii/\heii\ flux ratio, coupled with the photoionization-based estimates in this work, puts the column density required for \rfe{} to be $\gtrsim 10^{24}\,\rm{cm^{-2}}$. More recent works suggest a slightly higher value of these line ratios, for example, \aliii/\heii\ = 5.35$\pm$2.728 if the $\lambda$1900\AA~ blend is fitted with a combination of blue-shifted component that is characteristic for the low-density high-ionization outflowing component, and, a broad component that is typical for the high-density low-ionization part of the BLR \citep[][Paola Marziani, priv. comm.]{negreteetal12}. Certainly, a higher S/N ratio is needed to properly account for the issues mentioned above. An increased availability of optical-UV and NIR spectroscopic measurements, especially with the advent of the upcoming ground-based 10-metre-class \citep[e.g. Maunakea Spectroscopic Explorer,][]{2019BAAS...51g.126M} and 40 metre-class \citep[e.g. The European Extremely Large Telescope,][]{2015arXiv150104726E} telescopes; and space-based missions such as the James Webb Space Telescope and the Nancy Grace Roman Space Telescope would certainly be a welcome addition to help break this degeneracy.

On the other hand, for the cloud sizes, \citet{2009ApJ...707L..82F} find that the minimum column density required is $\sim$10$^{23}$ cm$^{-2}$ for gravity to overpower radiation pressure and allow infall of clouds as found by \citet{hu08}. Using arguments based on virial determinations of the black hole mass in AGNs by \citet{2008ApJ...678..693M}, \citet{2009ApJ...695..793N} also concludes that the column densities must substantially exceed $\sim$10$^{23}$ cm$^{-2}$ to avoid excessive effects of radiation pressure on the orbital velocities of the BLR clouds. Thus, there may be limited freedom to vary the column density to produce the wide range of optical \feii{} strength observed which then restricts the parameter space within $\lesssim$ 2 dexes in column density without accounting for significant electron scattering effects that start to become important at higher optical depths. Thus, with such constraints on the column densities and from Figure \ref{fig:Z-N}, we expect metallicities no greater than $\sim$30\zsun{} but still $\gtrsim$ 5\zsun{} to efficiently produce the required \rfe{} values in this case. From the point of view of recent advances in observations, only very recently we are starting to resolve the inner parsec scales in nearby AGNs using interferometric techniques \citep[]{sturm2018,2020arXiv200908463G} but mapping individual BLR clouds is something that remains elusive.
\section{Conclusions}
\label{sec:conclusion}

In this article, we examine the \rfe{} and \rcat{} emission in the broad-line regions of active galaxies. We probe the parameter space in terms of (a) ionization parameter, (b) the BLR cloud density, (c) the metal content in the BLR cloud, and (d) the size of the BLR cloud in terms of the cloud column density. We incorporate the observational broad-band SED of the prototypical Narrow-line Seyfert 1 galaxy, \textmyfont{I Zw 1}, that serves as the incident continuum that photoionizes the BLR cloud. In our previous paper (P20) and the first attempt of this paper, we are successful in reproducing the respective flux ratios (\rfe{} and \rcat{}) although it was noticed that the pairs of \un{} that correspond to the best estimates for the flux ratios do not recover reasonable line EWs for these low-ionization lines, including \hb{}. We evaluate the EWs for the entire grid of models and optimize the line EWs within reasonable covering factors (between 30\%-60\%) and recover also the flux ratios obtained from two epochs of prior observations for this source \citep{persson1988,murilo2016}. These new solutions are found to have ionization parameters (log \u{} $\sim$ -3.5) that are $\sim$ 2 dexes smaller than previous results. Although, the local cloud density remains nearly unchanged (log \n{} $\sim$ 12 cm$^{-3}$). This points towards a significant reduction in the flux that is incident on the BLR cloud than assumed before. We achieve this reduction in the flux by scaling the radial distance of this LIL emitting region obtained from our photoionization modelling to the radial distance obtained from the reverberation mapping estimate for \textmyfont{I Zw 1} using the standard radius-luminosity relation (R$_{\rm{BLR}} \propto \rm{L^{0.5}}$). This suggests that the broad-line region cloud doesn't see the same continuum seen by a distant observer that is emanated from the accretion disk, rather it sees a filtered, colder continuum that implies a lowering in the number of line-ionizing photons that irradiate the BLR. This in turn suggests smaller radial sizes than predicted earlier by photoionization modelling. This screening of the accretion disk continuum can be tied to the flux as a function of the cos\textit{i}, where \textit{i} is the inclination angle to the symmetry axis. Additionally, there can be a lowering of the photon flux due to anisotropy in the disk structure causing self-shadowing effects as the accretion rate increases in addition to light-bending effects. Our results are applicable to Type-1 Narrow-line Seyfert galaxies with high \feii{} emission, i.e. \textmyfont{I Zw 1}-like sources.

Independently from this aspect, our study still finds that to account for the adequate \rfe{} emission, the BLR needs to be selectively overabundant in iron. This is suggested by the requirement of higher than solar metallicities (Z$\gtrsim$10\zsun{}) to optimize the emission of optical \feii{}. On the other hand, the \rcat{} emission spans a broader range in metallicity, from solar to super-solar metallicities. In all these models the BLR cloud density is found to be consistent with our conclusions from prior works, i.e. \n{} $\sim$ 10$^{12}$ cm$^{-3}$ is required for the sufficient emission of \feii{} and \cat{}. We further our modelling to test and confirm the co-dependence between the metallicity and the cloud column density for these two ions and constrain the effective cloud sizes using metallicity constraints from UV line ratios shown to be effective tracers of the metal content in the BLR. Finally, we test the effect of inclusion of a turbulent velocity within the BLR cloud which informs us that the \rfe{} emission is positively affected by the inclusion of the microturbulence parameter. An interesting result obtained here is, when the microturbulence is invoked, there is a reduction in the value of the metallicity required to obtain optimal \rfe{} estimates, suggesting that microturbulence can act as a metallicity controller for the \feii{}. On the contrary, the \rcat{} cases are rather unaffected by the effect of microturbulence.


\begin{acknowledgements}
I thank the anonymous referee for useful suggestions that helped to improve the paper. I would like to thank Prof. Paola Marziani for performing a re-analysis of the \textmyfont{I Zw 1} spectrum to estimate the metallicity and Prof. Jian-Min Wang for providing his slim disk SED models for this project. I'd like to thank Prof. Bo\.zena Czerny, Prof. Paola Marziani, Dr Mary Loli Mart\'inez-Aldama and Dr Deepika Ananda Bollimpalli for fruitful discussions leading to the current state of the paper. I thank Prof. Bo\.zena Czerny, Prof. Paola Marziani and Mr Sushanta Kumar Panda for proof-reading the manuscript and suggesting corrections to improve the overall readability. The project was partially supported by the Polish Funding Agency National Science Centre, project 2017/26/\-A/ST9/\-00756 (MAESTRO  9) and MNiSW grant DIR/WK/2018/12.
\end{acknowledgements}

\section*{Softwares}
\textmyfont{CLOUDY} v17.02 (\citealt{f17}); \textmyfont{MATPLOTLIB}  (\citealt{hunter07}); \textmyfont{NUMPY} (\citealt{numpy}); \textmyfont{R} (\citealt{r_stats})

\bibliography{main}
\bibliographystyle{aa}

\end{document}